\documentclass[aps,10pt,pra,twocolumn,amsmath,amssymb,superscriptaddress,groupedaddress]{revtex4-1}

\usepackage[english]{babel} 
\usepackage[utf8]{inputenc}
\usepackage[T1]{fontenc}
\usepackage{graphicx}
\usepackage{epsfig}
\usepackage{graphicx}
\usepackage{subfigure}
\usepackage{qcircuit}
\usepackage{upgreek}
\usepackage{xcolor}

\usepackage[normalem]{ulem}

\usepackage{hyperref}
\hypersetup{
	colorlinks=true,
	linkcolor=red,			
	citecolor=green,		
	filecolor=magenta,		
	urlcolor=magenta		
}

\newcommand{\ket}[1]{\left\vert #1 \right\rangle}
\newcommand{\bra}[1]{\left\langle #1 \right\vert}
\newcommand{\abs}[1]{\left\vert #1 \right\vert}

\newcommand{\komm}[2]{\left[ #1 , #2 \right]}
\newcommand{\akomm}[2]{\left\{ #1 , #2 \right\}}

\renewcommand{\d}{\mathrm{d}}
\newcommand{\comma}{~,}
\newcommand{\fullstop}{~.}

\newcommand{\hc}{\mathrm{H.c.}}
\newcommand{\ie}{\emph{i.e.}}

\def\ValueFigIIabackend{ibmqx2}

\def\ValueFigIIaqbtpe{2}
\def\ValueFigIIaqbtme{4}
\def\ValueFigIIaDelta{0}

\def\ValueFigIIadt{0.1}

\def\ValueFigIIaAbstpe{1}
\def\ValueFigIIaAngletpe{5}

\def\ValueFigIIaAbstme{0.5}
\def\ValueFigIIaAngletme{-}
\def\ValueFigIIaAbstmp{0}

\def\ValueFigIIaPhaseUnit{\pi i/6}

\def\ValueFigIIbcbackend{ibmqx2}

\def\ValueFigIIbcqbtpe{2}
\def\ValueFigIIbcqbtme{2}
\def\ValueFigIIbcDelta{0}
\def\ValueFigIIbcepsilon{0.25}
\def\ValueFigIIbcdt{0.2}

\def\ValueFigIIbcAbstpe{2}
\def\ValueFigIIbcAngletpe{-}
\def\ValueFigIIbcGammame{1}
\def\ValueFigIIbcAbstme{1}
\def\ValueFigIIbcAngletme{2}
\def\ValueFigIIbcAbstmp{0}

\def\ValueFigIIbcPhaseUnit{\pi i/6}

\def\ValueFigIIIabackend{ibmqx2}

\def\ValueFigIIIaqbtpe{2}
\def\ValueFigIIIaqbtme{2}

\def\ValueFigIIIaepsilon{0.25}
\def\ValueFigIIIadt{0.2}
\def\ValueFigIIIaNtrot{3}

\def\ValueFigIIIaAbstpe{2}
\def\ValueFigIIIaAngletpe{-}
\def\ValueFigIIIaGammame{1}
\def\ValueFigIIIaAbstme{2}
\def\ValueFigIIIaAngletme{2}
\def\ValueFigIIIaAbstmp{0}

\def\ValueFigIIIaPhaseUnit{\pi i/6}

\def\ValueFigIIIbDelta{0}

\def\ValueFigIIIcDelta{0}
\def\ValueFigIIIcepsilon{0.25}

\def\ValueFigIIIdepsilon{0.25}
\def\ValueFigIIIddt{0.2}

\def\ValueFigIIIdAbstpe{2}
\def\ValueFigIIIdAngletpe{-2}
\def\ValueFigIIIdGammame{1.25}
\def\ValueFigIIIdAbstme{2}
\def\ValueFigIIIdAngletme{-2}
\def\ValueFigIIIdAbstmp{0}

\def\ValueFigIIIdPhaseUnit{\pi i/6}

\begin{document}

\selectlanguage{english}

\title{Quantum Synchronization on the IBM Q System}

\author{Martin Koppenh\"ofer}
\affiliation{Department of Physics, University of Basel, Klingelbergstrasse 82, CH-4056 Basel, Switzerland}

\author{Christoph Bruder}
\affiliation{Department of Physics, University of Basel, Klingelbergstrasse 82, CH-4056 Basel, Switzerland}

\author{Alexandre Roulet}
\affiliation{Department of Physics, University of Basel, Klingelbergstrasse 82, CH-4056 Basel, Switzerland}

\date{\today}

\begin{abstract}
	We report the first experimental demonstration of quantum synchronization. 
	This is achieved by performing a digital simulation of a single spin-$1$ limit-cycle oscillator on the quantum computers of the IBM Q System.
	Applying an external signal to the oscillator, we verify typical features of quantum synchronization and demonstrate an interference-based quantum synchronization blockade. 
	Our results show that state-of-the-art noisy intermediate-scale quantum computers are powerful enough to implement realistic dissipative quantum systems. Finally, we discuss limitations of current quantum hardware and define requirements necessary to investigate more complex problems.
\end{abstract}

\maketitle

\section{Introduction}
Synchronization, \ie, the adjustment of the rhythm of a self-sustained oscillation to a weak perturbation, is a universal feature of many complex dynamical systems \cite{Pikovsky}.
Classical synchronization has been demonstrated in a variety of very different setups ranging from electrical circuits to biological neuron systems \cite{Adler-IRE.1697085,Pecora-PhysRevLett.64.821,Chagnac-JNeurophysiol.62.1149}. 
Several proposals have been made to study quantum effects of synchronization in 
	superconducting circuits \cite{Zhirov-EPJD.38.375,Nigg-PhysRevA.97.013811}, 
	optomechanical systems \cite{Ludwig-PhysRevLett.111.073603,Walter-PRL.112.094102}, 
	trapped ions \cite{Lee-PRL.111.234101,Hush-PhysRevA.91.061401}, and
	nanomechanical oscillators \cite{Holmes-PhysRevE.85.066203}.
However, all the experimental demonstrations of synchronization reported to date on these platforms were operating in the classical regime \cite{Zalalutdinov-ApplPhysLett.83.3281,Hossein-Zadeh-ApplPhysLett.93.191115,Zhang-PhysRevLett.115.163902,Bagheri-PhysRevLett.111.213902,Shlomi-PhysRevE.91.032910,Seitner-PhysRevLett.118.254301,Gil-Santos-PhysRevLett.118.063605,Bekker-optica.4.1196,Toth-PhysLettA.382.2233,Huang-OptExpr.26.8275}, because of the challenge of sustaining a highly nonlinear oscillator in the quantum regime.

Here, we report the first experimental demonstration of quantum synchronization. 
Our quantum limit-cycle oscillator is implemented in a single spin-$1$ system, which was recently introduced as the smallest possible system that can be synchronized \cite{Roulet-PRL.121.053601}.
We use two qubits of a quantum computer to implement the desired spin-$1$ system while the remaining qubits play the role of the environment sustaining the oscillation. 
The advantage of this approach is that the nonlinear dissipation required to study quantum synchronization corresponds to easily engineered single-qubit relaxation, which enables the study of nonlinear oscillators in the quantum regime.
With this mapping in place, we perform a digital quantum simulation \cite{NielsenChuang,Lloyd-science.273.1073} of spin-$1$ synchronization dynamics on the publicly available few-qubit quantum computers at the IBM Q System \cite{IBMQX}. 
More specifically, we program the universal quantum computer such that it approximates the time evolution of the spin-$1$ system of interest and we extract the state of the spin-$1$ system by measuring the two qubits.
In this sense, the two qubits encoding the spin-$1$ system represent an experimental realization of a spin-$1$ limit-cycle oscillator.

The ongoing efforts to build a quantum computer have resulted in noisy intermediate-scale quantum (NISQ) devices, which are constantly improving in terms of decoherence and relaxation times, gate fidelities, and readout fidelities \cite{Preskill-Quantum.2.2521}. 
NISQ devices have become a highly relevant platform for simulating realistic physical problems and they have already been used to find quantum ground states \cite{Peruzzo-ncomms.5.4213,Kandala-Nature.549.242,Reiner-QST.4.035005} and to simulate closed-system quantum many-body dynamics \cite{Pollmann-1906.06343}. 
Moreover, it has been shown that they can in principle be used to simulate the dynamics of dissipative quantum systems \cite{Lloyd-PhysRevA.65.010101,Bacon-PhysRevA.64.062302,Kliesch-PhysRevLett.107.120501,Garcia-Perez-1906.07099}. 
Our results demonstrate that state-of-the-art NISQ devices are indeed able to study complex dissipative quantum systems that were not realized experimentally before.

\section{System and mapping}
\label{sec:System}
\begin{figure*}
	\centering
	\includegraphics[width=\textwidth]{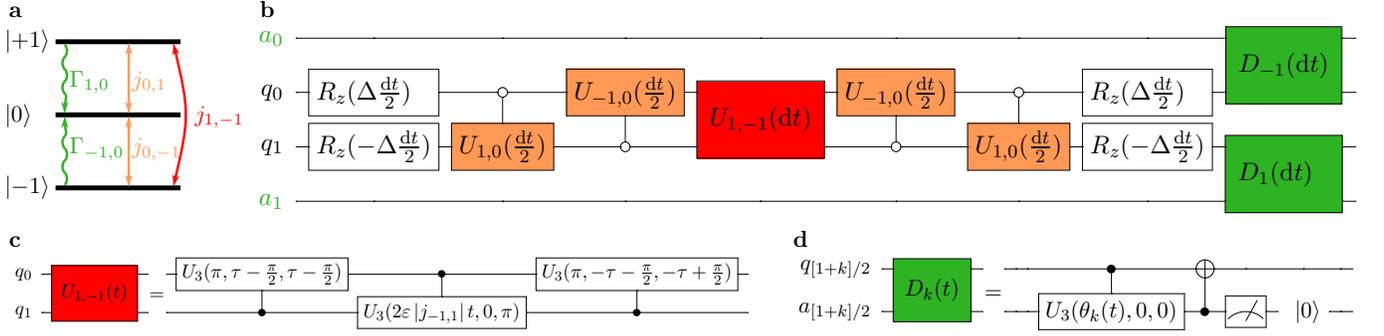}
	\caption{
	(a) Energy level diagram of a spin-$1$ system hosting a limit-cycle oscillator. 
		The limit cycle is stabilized by dissipative transitions towards the state $\ket{0}$ at rates $\Gamma_{\pm 1,0}$ and is subjected to an external signal which drives transitions $j_{k,l}$ between the spin-$1$ states.
	(b) Quantum-circuit implementation of the synchronization dynamics for a timestep $\d t$, obtained by a Suzuki-Trotter decomposition.
		The gates shown in white correspond to the free evolution of the oscillator while the other circuit components correspond to the transitions of the same color in (a).
		Here, $R_z(\theta) = U_3(0,0,\theta)$ is the phase gate and the signals $j_{0,\pm 1}$ are mapped onto controlled gates 
		$U_{\pm 1,0}(t) = U_3(-2 \varepsilon \abs{j_{0,\pm 1}} t,\arg(j_{0,\pm 1}) - \frac{3\pi}{2},-\arg(j_{0,\pm 1}) - \frac{\pi}{2})$. 
		The $U_3(\theta, \varphi, \lambda)$ gate is a basis gate of the IBM quantum computer, defined in Eq.~\eqref{eqn:SM:U3}.
		Open (solid) circles indicate a controlled gate conditioned on the control qubit $q_k$ being in $\ket{0}_{q_k}$ ($\ket{1}_{q_k}$), see Eq.~\eqref{eqn:Mapping:Representation}.
	(c) Trotter step of the $j_{1,-1}$ signal using three controlled $U_3(\theta, \varphi, \lambda)$ gates, where $\tau= \arg(j_{-1,1})$. 
	(d) Implementation of relaxation dynamics with $\theta_k(t) = 2 \arcsin (\sqrt{\Gamma_{k,0} t})$.
		Note that the two dissipative steps in (b) could also be applied sequentially to a single ancillary qubit.
	}
	\label{fig:Mapping}
\end{figure*}

We consider the synchronization of a single spin-$1$ limit-cycle oscillator to an external signal of strength $\varepsilon$ that is described by a Hamiltonian $\hat{H}_\mathrm{signal}$. 
The dynamics in a frame rotating at the signal frequency and under a rotating wave approximation is given by the quantum master equation ($\hbar = 1$) \cite{Koppenhoefer-PhysRevA.99.043804}
\begin{align}
	\frac{\d}{\d t} \hat{\rho} &= -i \komm{\Delta \hat{S}_z + \varepsilon \hat{H}_\mathrm{signal}}{\hat{\rho}} \nonumber \\
	&\hphantom{=}\ + \Gamma_\mathrm{-1,0} \mathcal{D}[\hat{S}_+ \hat{S}_z] \hat{\rho} + \Gamma_{1,0} \mathcal{D}[\hat{S}_- \hat{S}_z] \hat{\rho} \fullstop
	\label{eqn:Mapping:QME}
\end{align}
Here, $\hat{S}_z$ is the spin-$1$ operator along the quantization axis and $\Delta = \omega_0 -\omega_\mathrm{signal}$ is the detuning between the spin precession frequency $\omega_0$ and the signal frequency $\omega_\mathrm{signal}$. 
By $\hat{S}_\pm$ we denote the spin raising and lowering operators, $\Gamma_{-1,0}$ and $\Gamma_{1,0}$ are the decay rates towards the state $\ket{0}$, and $\mathcal{D}[\hat{O}]\hat{\rho} = \hat{O} \hat{\rho} \hat{O}^\dagger - \frac{1}{2} \akomm{\hat{O}^\dagger \hat{O}}{\hat{\rho}}$ is a Lindblad dissipator. 
The signal Hamiltonian is given by $\hat{H}_\mathrm{signal} = j_{0,1} \hat{S}_z \hat{S}_+/\sqrt{2} - j_{0,-1} \hat{S}_z \hat{S}_-/\sqrt{2}  + j_{-1,1} \hat{S}_+^2/2 + \hc$ where the complex coefficients $j_{k,l}$ determine the relative amplitude and phase of the three possible transitions in a spin-$1$ system, as sketched in Fig.~\ref{fig:Mapping}(a). 
For instance, the combination $j_{0,1} = j_{0,-1}^*$ and $j_{-1,1} = 0$ corresponds to a semiclassical signal, while $j_{0,1} = j_{0,-1} = 0$ and $j_{-1,1} \neq 0$ corresponds to a squeezing signal.

To simulate a quantum system on a quantum computer, its Hilbert space $\mathcal{H}_\mathrm{sys}$ needs to be mapped onto the logical Hilbert space $\mathcal{H}_\mathrm{qc}$ of the quantum computer. 
We choose to represent the three spin-$1$ states in terms of the following two-qubit states. 
\begin{align}
	\ket{+1} &= \ket{1}_{q_1} \otimes \ket{0}_{q_0} \comma \nonumber \\
	\ket{0} &= \ket{0}_{q_1} \otimes \ket{0}_{q_0} \comma \nonumber \\
	\ket{-1} &= \ket{0}_{q_1} \otimes \ket{1}_{q_0} \fullstop 
	\label{eqn:Mapping:Representation}
\end{align}
Note that this encoding gives rise to a fourth state $\ket{X} = \ket{1}_{q_1} \otimes \ket{1}_{q_0}$ outside the spin-1 Hilbert space, which needs to be isolated from the other states.

Next, the system's continuous dynamics~\eqref{eqn:Mapping:QME} has to be translated to the level of logical qubits, to which we can only apply a finite set of discrete unitary gates. 
The exact time evolution is approximated by a series of many transformations that propagate the system's state for a small timestep $\d t$. 
For the unitary part of Eq.~\eqref{eqn:Mapping:QME}, this is achieved by means of a Suzuki-Trotter decomposition \cite{NielsenChuang}. 
Simulating the remaining non-unitary dissipative dynamics may seem challenging given that we can only apply unitary gates.
However, this task can be achieved by simulating discrete-time unitary dynamics on an extended system where ancillary degrees of freedom mimic a dissipative environment.
In fact, it has been shown that this environment can even be modeled by a single resettable qubit \cite{Lloyd-PhysRevA.65.010101}.

In our case, a single Trotter time step $\d t$ that approximates the dynamics~\eqref{eqn:Mapping:QME} up to corrections of the order $\d t^3$ is shown in Figs.~\ref{fig:Mapping}(b-d).
This is one of our main results. 
The signal Hamiltonian $\hat{H}_\mathrm{signal}$ is implemented by controlled two-qubit rotations such that the undesired state $\ket{X}$ remains decoupled from the spin-$1$ system. 
Our mapping~\eqref{eqn:Mapping:Representation} has the benefit that the limit-cycle state $\ket{0}$ corresponds to the ground state $\ket{0}_{q_1} \otimes \ket{0}_{q_0}$ of the qubits.
Thus, the dissipative stabilization of the limit cycle translates to energy relaxation processes on the two qubits $q_0$ and $q_1$.
This allows us to implement the required nonlinear dissipation in the quantum regime with minimal complexity:
The non-unitary circuit $D_k$ performing a measurement and subsequent reset of the ancilla qubit, shown in Fig.~\ref{fig:Mapping}(d), implements single-qubit relaxation with a tunable relaxation rate $\Gamma_{k,0}$ \cite{NielsenChuang}. 
As discussed in App.~\ref{sec:App:Dissipation}, we are effectively implementing a photon-counting quantum-trajectory simulation of the quantum master equation~\eqref{eqn:Mapping:QME} granted that the condition $\Gamma_{k,0} \d t \ll 1$ holds. 
Each experimental run of the circuit calculates a random quantum trajectory of a pure state and the dynamics of $\hat{\rho}$ can be recovered by an ensemble average over many quantum trajectories \cite{WisemanMilburn}.

\section{Methods}
\label{sec:Methods}
The data presented in this article have been collected on the publicly accessible NISQ computer \textsc{ibmqx2} between September 30 and October 7, 2019.
This quantum computer provides $5$ qubits in a star-shaped geometry where CNOT operations can be performed between the central qubit $2$ and all other qubits $0$, $1$, $3$, and $4$ \cite{IBMHardware}. 
Additional CNOT operations are provided between the qubits $0$ and $1$ as well as $3$ and $4$, which we do not use, however. 
The maximum CNOT error rate is below $2\,\%$ and the central qubit $2$ has a $T_2$ time of approximately $70\,\upmu \mathrm{s}$.

We used the python API \textsc{Qiskit}~\cite{Qiskit} to define quantum circuits, to submit them to the quantum computer, and to evaluate the measurement results returned from the quantum computer.
Before submission, each circuit has been mapped (\emph{transpiled}) to a set of basis gates of the IBM devices, which are a two-qubit $\mathrm{CNOT}$ gate, the single-qubit $U_3(\theta, \varphi, \lambda)$ gate defined by
\begin{align}
	U_3 \ket{0}_{q_j} &= \cos \frac{\theta}{2} \ket{0}_{q_j} + e^{i \varphi} \sin \frac{\theta}{2} \ket{1}_{q_j} \comma 
	\label{eqn:SM:U3}\\
	U_3 \ket{1}_{q_j} &= -  e^{i \lambda} \sin \frac{\theta}{2} \ket{0}_{q_j} + e^{i \lambda + i \varphi} \cos \frac{\theta}{2} \ket{1}_{q_j} \comma \nonumber
\end{align}
and the single-qubit gates $U_2(\varphi,\lambda) = U_3(\pi/2,\varphi,\lambda)$ and $U_1(\lambda) = U_3(0,0,\lambda)$. 
The final state has been reconstructed by a quantum state tomography using the builtin \textsc{Qiskit} functions implementing Ref.~\onlinecite{Smolin-PhysRevLett.108.070502}.
Each quantum circuit has been executed with the maximum possible number of $8192$ repetitions.

The queuing system of the quantum computer allows one to group several quantum circuits to batch jobs, which are treated as a single task such that all quantum circuits in the batch job are executed successively.
We grouped quantum circuits generating the time evolution for different numbers of time steps or for scans of different values of a parameter. 
At the beginning of each batch job, two calibration circuits have been added to measure the readout error of the qubits $q_0$ and $q_1$.
The readout error of the central qubit $2$ is approximately $1\,\%$ \cite{IBMQX}.
Based on these calibration results, the measurement errors of all subsequent measurements have been mitigated using \textsc{Qiskit} methods. 
To validate the stability of the error mitigation procedure and to rule out drifts of the device parameters during data collection, each batch job has been submitted three times.
The corresponding standard deviation is indicated by the error bars in the plots, which are smaller than the plot markers. 
Note that these error bars capture only statistical measurement errors and the short-term stability of the device parameters on a timescale of hours. 
Since the parameters of the quantum computer vary on a timescale of days, the quantum computers are recalibrated on a daily basis. 
Therefore, numerical changes of the results obtained for small signal strength $\varepsilon \to 0$ are expected if data obtained on different days is compared.

\section{Device characterization}
\label{sec:Characterization}
By iteratively applying $N$ Trotter steps on an ideal quantum computer, an initial state is evolved to a final state at time $T = N \d t$. 
In a first step, we assess whether this is the case on an actual NISQ device by testing the elements of the decomposition shown in Figs.~\ref{fig:Mapping}(b-d). 
We also discuss the restrictions imposed by the limited capabilities of state-of-the-art quantum computers.

\begin{figure*}	
	\centering
	\includegraphics[width=\textwidth]{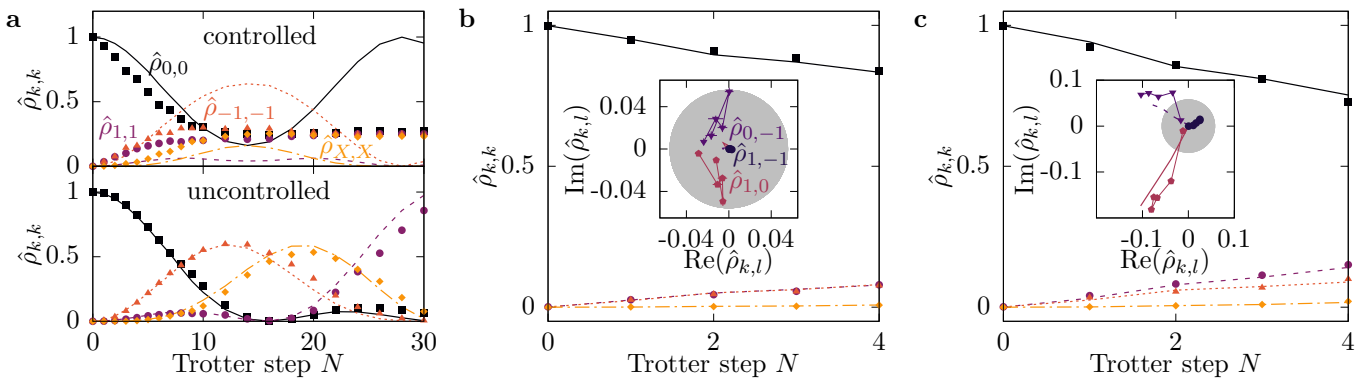}
	\caption{
	(a) Time evolution of the state $\ket{0}$ under the semiclassical signal components $j_{0,\pm 1}$ (markers) on a noisy intermediate-scale quantum device (NISQ) and ideal noise-free time evolution (lines). 
		The upper plot is obtained for \emph{controlled} $U_{\pm 1,0}$ gates as shown in the circuit diagram in Fig.~\ref{fig:Mapping}(b) for 
		$\varepsilon \d t = \ValueFigIIadt$, $j_{-1,0} = \ValueFigIIaAbstme \times e^{\ValueFigIIaAngletme  \ValueFigIIaPhaseUnit}$, $j_{1,0} = \ValueFigIIaAbstpe \times e^{\ValueFigIIaAngletpe \ValueFigIIaPhaseUnit}$, $j_{-1,1} = \ValueFigIIaAbstmp$, and $\Delta/\varepsilon = \ValueFigIIaDelta$. 
		In the lower plot, \emph{uncontrolled} single-qubit $U_{\pm 1,0}$ gates were used with the same parameters.
		The data has been collected on the \textsc{\ValueFigIIabackend} computer on qubits $q_0 = \ValueFigIIaqbtme$ and $q_1 = \ValueFigIIaqbtpe$.
	(b) Dissipative stabilization of the limit-cycle state $\ket{0}$ if no signal is applied, $j_{\pm 1,0} = j_{-1,1} = \ValueFigIIbcDelta$, on a NISQ device (markers) and theoretical expectation taking into account noise (lines).
		The dynamics of the coherences in the inset is illustrated by the thin connecting lines.
		The gray circle defines the level of the noise due to the dissipative limit-cycle stabilization.
		Parameters: 
		$\Gamma_{1,0} \d t = \ValueFigIIbcdt$, $\Delta/\Gamma_{1,0} = \ValueFigIIbcDelta$, $\varepsilon/\Gamma_{1,0} = \ValueFigIIbcepsilon$, and $\Gamma_{-1,0}/\Gamma_{1,0} = \ValueFigIIbcGammame$. 
		Data has been collected on \textsc{\ValueFigIIbcbackend} on qubits $q_0 = \ValueFigIIbcqbtme$ and $q_1 = \ValueFigIIbcqbtpe$ in sequential runs.
	(c) Demonstration of the onset of synchronization if both the signal and the dissipative stabilization of the limit-cycle state are switched on, $j_{-1,0} = \ValueFigIIbcAbstme \times e^{\ValueFigIIbcAngletme \ValueFigIIbcPhaseUnit}$, $j_{1,0} = \ValueFigIIbcAbstpe \times e^{\ValueFigIIbcAngletpe \ValueFigIIbcPhaseUnit}$, and $j_{-1,1} = \ValueFigIIbcAbstmp$. 
		The signal builds up coherences beyond the noise level of the limit cycle.
		The data has been averaged over three runs, each having 8192 repetitions per circuit. 
		The corresponding error bars are smaller than the plot markers.
	}
	\label{fig:Characterization}
\end{figure*}

Figure~\ref{fig:Characterization}(a) shows the time evolution of the initial state $\ket{0}$ under the signal components $j_{0,\pm 1}$ on a NISQ device and the corresponding ideal noise-free result.
Simulations of the exact dynamics, given by Eq.~\eqref{eqn:Mapping:QME}, have been performed using the python package \textsc{Qutip} \cite{Qutip}. 
Controlled two-qubit gates are found to induce strong depolarization errors that evolve the initial state $\ket{0}$ to a completely mixed state after only a few Trotter steps. 
This result is also confirmed by simulations taking into account a noise model of the IBM quantum computers provided in the python API \textsc{Qiskit}. 
Given that already the signal component suffers from severe depolarization errors, it is not feasible to perform the time evolution as shown in Figs.~\ref{fig:Mapping}(b-d) on a NISQ device. 
However, in the synchronization regime, most of the population remains in the limit-cycle state $\ket{0}$. 
Therefore, it is possible to circumvent the problem of depolarizing errors by considering a modified circuit consisting only of uncontrolled single-qubit $U_{\pm 1,0}$ rotations. 
This is discussed in more detail in the next section. 
The single-qubit-rotation error rates on the IBM quantum computers are about an order of magnitude smaller than the two-qubit controlled-not (CNOT) error rate \cite{IBMQX}.
Consequently, the uncontrolled implementation of the signal using only single-qubit rotations reproduces the ideal noise-free result almost perfectly over a much larger range of Trotter steps, as shown in Fig.~\ref{fig:Characterization}(a).

Figure~\ref{fig:Characterization}(b) demonstrates the dissipative stablization of the limit cycle state $\ket{0}$ if no signal is applied, $j_{\pm 1,0} = j_{1,-1} = 0$.
Once more, the controlled two-qubit operations contained in the operations $D_k$ induce a decay of the state $\ket{0}$ towards a completely mixed state. 
Surprisingly, the noise induced by the dissipative stabilization is such that the limit-cycle state shows a small amount of coherence.
This effect is not captured by the simple noise model provided in \textsc{Qiskit}.
The corresponding results demonstrating that the initial states $\ket{\pm 1}$ evolve to the limit-cycle state $\ket{0}$ under the action of the dissipative terms $D_k$ are given in App.~\ref{sec:App:Stabilization}.
There, we also discuss an alternative implementation of the dissipative stabilization that requires fewer two-qubit gates and can be used to minimize noise in the coherences.

Besides the strong depolarizing effect of two-qubit gates, another limitation of IBM's current quantum computers is that they do not allow measurement and reset operations of qubits in the middle of a quantum circuit. 
This means that we must use a new ancillary qubit in each timestep and measure all of them at the end of the time evolution. 
Therefore, the maximum number of Trotter steps we can apply is bounded by the number of available ancillary qubits on a quantum computer. 
Moreover, since SWAP operations are composed of three CNOT gates and suffer strong depolarizing errors, we can only use qubits that are directly connected to the system qubit $q_j$ to be relaxed, which limits us to at most four timesteps.
At the moment, this is the most severe limitation for the simulation of dissipative quantum systems on the device. 
We expect that it will be lifted in the near future.

\section{Dealing with hardware constraints}
\label{sec:Dealing}
The paradigm of quantum synchronization allows us to adapt the quantum circuit shown in Fig.~\ref{fig:Mapping} to the limitations of IBM's quantum computers.
Specifically, the signal strength is linearly proportional to a small dimensionless parameter $0 \leq \eta \ll 1$ that ensures that $\hat{H}_\mathrm{signal}$ is only a small perturbation to the limit-cycle state \cite{Koppenhoefer-PhysRevA.99.043804}. 
Thus, the amplitudes of the coherences $\hat{\rho}_{\pm 1,0}$ are of order $\eta$ and the populations of the states $\ket{\pm 1}$ are of order $\eta^2$.
That is, they are strongly suppressed as compared to the limit-cycle state $\ket{0}$ having a population of $\mathcal{O}(1)$. 
Under these conditions, we can replace the controlled two-qubit gates $U_{\pm 1,0}$ by uncontrolled single-qubit rotations.
In principle, the signal will now build up coherences $\hat{\rho}_{k,X}$ between the spin-$1$ states and the state $\ket{X}$ and it will transfer population to the state $\ket{X}$. 
However, both effects can be safely ignored, in particular on a noisy system, because the coherences $\hat{\rho}_{k,X}$ and the population $\hat{\rho}_{X,X}$ are only of order $\eta^3$ and $\eta^4$, respectively.
Moreover, since the relaxation mechanism $D_k$ takes the state $\ket{X}$ back to $\ket{\pm 1}$, there is no risk to trap population in $\ket{X}$. 
Plots verifying that the coherences $\hat{\rho}_{k,X}$ are well below the limit-cycle noise threshold are shown in App.~\ref{sec:App:Stabilization}.

Having replaced controlled by uncontrolled rotations, if we additionally restrict ourselves to semiclassical signals, \ie, $j_{-1,1} = 0$, the entire unitary part of the time evolution~\eqref{eqn:Mapping:QME} can be simulated using only single-qubit rotations. 
The qubits $q_0$ and $q_1$ can now be independently assigned to physical qubits of the quantum computer, which allows us to 
evolve the qubits $q_{0}$ and $q_{1}$ sequentially in two consecutive runs on the $5$-qubit \textsc{ibmqx2} quantum computer.
The full spin-$1$ density matrix is reconstructed from quantum state tomographies of the qubits $q_0$ and $q_1$ at the end of each time evolution.

Given the fixed connectivity and SWAP fidelities of IBM's current quantum computers, the limit on the available Trotter steps imposed by the device connectivity cannot be evaded. 
As a consequence, quantum simulation of the steady-state solution of Eq.~\eqref{eqn:Mapping:QME} is out of reach, but we are able to demonstrate the transient buildup of synchronization, as shown in Fig.~\ref{fig:Characterization}(c).

\section{Results}
\begin{figure*}	
	\includegraphics[width=\textwidth]{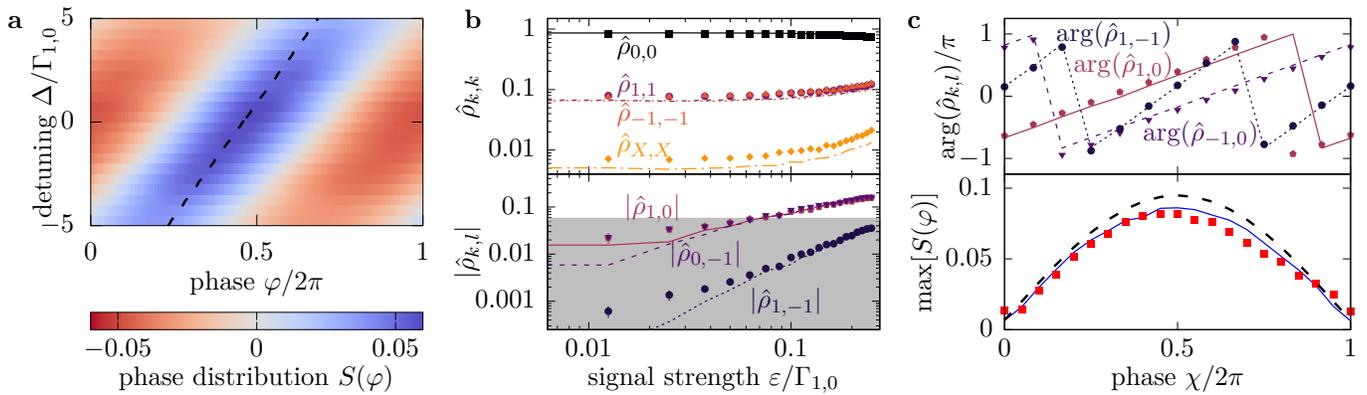}
	\centering
	\caption{
	(a) Phase distribution $S(\varphi)$ of the spin-$1$ limit-cycle oscillator as a function of the detuning $\Delta$ between its natural frequency and the signal frequency after $N = \ValueFigIIIaNtrot$ timesteps. 
		The dashed black line indicates the theoretical expectation of the position of the maximum of $S(\varphi)$, obtained by combining Eqs.~\eqref{eqn:Mapping:QME} and~\eqref{eqn:Results:S}.
		Parameters are 
		$\Gamma_\mathrm{-1,0}/\Gamma_\mathrm{1,0} = \ValueFigIIIaGammame$, 
		$\Gamma_\mathrm{1,0} \d t = \ValueFigIIIadt$, 
		$\varepsilon/\Gamma_\mathrm{1,0} = \ValueFigIIIaepsilon$, 
		$j_{-1,0} = \ValueFigIIIaAbstme \times e^{\ValueFigIIIaAngletme \ValueFigIIIaPhaseUnit}$, 
		$j_{1,0} = \ValueFigIIIaAbstpe \times e^{\ValueFigIIIaAngletpe \ValueFigIIIaPhaseUnit}$, and 
		$j_{-1,1} = \ValueFigIIIaAbstmp$. 
	(b) Populations and coherences as a function of the signal strength $\varepsilon$ for 
		$\Delta/\Gamma_{1,0} = \ValueFigIIIbDelta$.
		The gray background indicates the noise level of the coherences introduced in Fig.~\ref{fig:Characterization}(b).
	(c) Upper panel: 
		Phase of the coherences if the overall phase $\chi$ of the signals, $j_{\pm 1,0} e^{i \chi}$, is varied for 
		$\Delta/\Gamma_{1,0} = \ValueFigIIIcDelta$ and $\varepsilon/\Gamma_{1,0} = \ValueFigIIIcepsilon$.
		Lower panel: 
		Demonstration of an interference-based quantum synchronization blockade if the phase of only one of the signals is varied, $j_{-1,0} = e^{i \chi} \times \ValueFigIIIdAbstme \times e^{\ValueFigIIIdAngletme \ValueFigIIIdPhaseUnit}$ and $j_{1,0} = \ValueFigIIIdAbstpe \times e^{\ValueFigIIIdAngletpe \ValueFigIIIdPhaseUnit} = \mathrm{const}$.
		Data points are the result obtained on a NISQ device, the solid line corresponds to a simulation taking into account noise, and the dashed line describes the theory result.
		Parameters are 
		$\Gamma_\mathrm{-1,0}/\Gamma_{1,0} = \ValueFigIIIdGammame$, 
		$\Gamma_{1,0} \d t = \ValueFigIIIddt$, 
		$\varepsilon/\Gamma_{1,0} = \ValueFigIIIdepsilon$, and 
		$j_{-1,1} = \ValueFigIIIdAbstmp$. 
		All data of this figure has been collected on the \textsc{\ValueFigIIIabackend} computer on qubits $q_0 = \ValueFigIIIaqbtme$ and $q_1 = \ValueFigIIIaqbtpe$ in sequential runs.
	}
	\label{fig:Results}
\end{figure*}

We now demonstrate typical features of quantum synchronization on the IBM Q System \cite{IBMQX}. 
Figure~\ref{fig:Results}(a) shows the phase distribution of the limit-cycle oscillator, calculated from the experimentally obtained density matrix according to the analytical formula \cite{Koppenhoefer-PhysRevA.99.043804}
\begin{align}
	S(\varphi) 
	&= \frac{3}{8 \sqrt{2}} \abs{\hat{\rho}_{1,0} + \hat{\rho}_{0,-1}} \cos \left[ \varphi + \arg( \hat{\rho}_{1,0} + \hat{\rho}_{0,-1}) \right] \nonumber \\
	&+ \frac{1}{2 \pi} \abs{\hat{\rho}_{1,-1}} \cos \left[ 2 \varphi + \arg(\hat{\rho}_{1,-1}) \right] \comma
	\label{eqn:Results:S}
\end{align}
as a function of the signal detuning. 
The dashed black line indicates the expected position of the peak of $S(\varphi)$ according to Eq.~\eqref{eqn:Mapping:QME}. 
The small differences in the positions of the maximum stem from a detuning dependence of the limit cycle stabilization mechanism due to device imperfections.
Figure~\ref{fig:Results}(b) confirms that the magnitude of the coherences between the spin eigenstates grows linearly with the overall signal strength $\varepsilon$, whereas the populations change only quadratically in $\varepsilon$.
Therefore, the applied signal perturbs the limit-cycle state only weakly and we operate in the regime of synchronization.
The buildup of the coherence $\hat{\rho}_{-1,1}$ is due to higher-order effects and scales proportional to $\varepsilon^2$. 
Finally, a global phase applied to the signals, $j_{\pm 1,0} \to e^{i \chi} j_{\pm 1,0}$, rotates the phase of the coherences accordingly as demonstrated in the upper panel of Fig.~\ref{fig:Results}(c). 
By rotating only the phase of one of the signal components, \ie, $j_{-1,0} \to e^{i \chi} j_{-1,0}$ but $j_{1,0} = \mathrm{const}$, the coherences $\hat{\rho}_{1,0}$ and $\hat{\rho}_{0,-1}$ in Eq.~\eqref{eqn:Results:S} can be tuned to interfere destructively, which manifests itself in an interference-based quantum synchronization blockade \cite{Koppenhoefer-PhysRevA.99.043804} and is demonstrated in the lower panel of Fig.~\ref{fig:Results}(c). 
This result is the first experimental demonstration of quantum effects in synchronization.

\section{Conclusion}
Understanding dissipative quantum systems is of high relevance for quantum sensing \cite{Wiersig-PhysRevA.93.033809}, quantum information processing \cite{Metelmann-PhysRevX.5.021025}, and quantum state preparation \cite{Poyatos-PhysRevLett.77.4728}.
Simulating dissipative systems is much harder than simulating a comparable closed system since one has to account for environmental degrees of freedom.
For instance, even for a moderate network size of approximately $20$ limit-cycle oscillators, classical simulation approaches will fail.
Our results demonstrate for the first time that state-of-the-art NISQ devices enable the study of realistic dissipative quantum systems.
However, they also reveal that known approaches to simulate dissipative quantum systems~\cite{Lloyd-PhysRevA.65.010101,Bacon-PhysRevA.64.062302,Kliesch-PhysRevLett.107.120501,Garcia-Perez-1906.07099} face obstacles when applied to more complex dissipative quantum systems, since 
	(i) the two-qubit gate fidelities~\cite{IBMHardware} are at least an order of magnitude too low, 
	(ii) missing qubit reset operations complicate the quantum circuit, and 
	(iii) the effective connectivity of the device is too low. 
	In a network of dissipative quantum systems, exchange interactions built out of controlled gates, such as the $U_{1,-1}$ gate, will become indispensable.

Despite these hardware limitations, we were able to experimentally demonstrate for the first time the synchronization of a quantum limit-cycle oscillator by a digital quantum simulation on the IBM Q System.
We observed a purely quantum effect in synchronization, namely a quantum interference-based synchronization blockade. 
Thus state-of-the-art NISQ computers are a useful tool to study simple realistic dissipative quantum systems and our results will provide a guideline for the development of new quantum computers and novel algorithms enabling the study of dissipative quantum systems on current hardware.

\begin{acknowledgments}
We would like to thank M.\ Marthaler, N.\ L\"orch, J.\ Wootton, and M.\ Mergenthaler for fruitful discussions. 
This work was financially supported by the Swiss National Science Foundation (SNSF) and the NCCR Quantum Science and Technology.

\emph{Note added}: After completion of this work, we became aware of Ref.~\onlinecite{arxiv.1910.11832} which studies quantum synchronization effects in an ensemble of spin-$1$ $^{87}$Rb atoms. 
\end{acknowledgments}

\appendix

\section{Implementing single-qubit relaxation}
\label{sec:App:Dissipation}

\begin{figure}
	\centering
	\includegraphics[width=.48\textwidth]{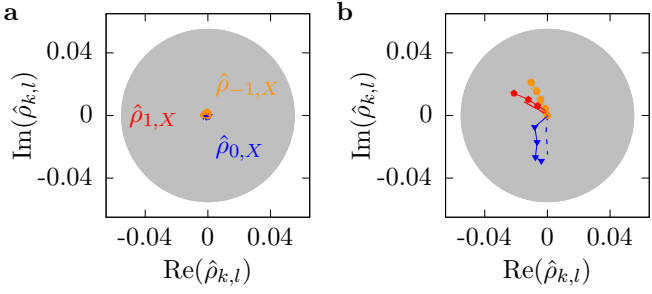}
	\caption{
	Time evolution of the coherences $\hat{\rho}_{1,X}$, $\hat{\rho}_{-1,X}$, and $\hat{\rho}_{0,X}$ for the parameters of Fig.~\ref{fig:Characterization}.
	(a) Only dissipative stabilization of the limit cycle is switched on, corresponding to Fig.~\ref{fig:Characterization}(b). 
	(b) An additional external signal is applied, corresponding to the Fig.~\ref{fig:Characterization}(c).
	The gray circle denotes the noise level of the limit-cycle state.
	}
	\label{fig:SM:FigA2a}
\end{figure}

As discussed in Sec.~\ref{sec:System}, our state representation~\eqref{eqn:Mapping:Representation} is chosen such that the dissipative stabilization of the limit-cycle state $\ket{0}$ in Eq.~\eqref{eqn:Mapping:QME} translates to a relaxation of the logical qubits towards the joint ground state $\ket{0}_{q_1} \otimes \ket{0}_{q_0}$. 
In principle, one could take advantage of the natural energy relaxation in the quantum computer to stabilize the limit cycle at the natural relaxation rate $\Gamma_\mathrm{rel}$. 
However, this is not sufficient if we want to study synchronization effects for the following reason.
An external signal $\hat{H}_\mathrm{signal}$ creates coherences between the spin-$1$ states at a certain rate $\Gamma_\mathrm{signal}$, which must be smaller than the rate $\Gamma_\mathrm{rel}$ at which the limit cycle is stabilized to satisfy the paradigm of synchronization \cite{Koppenhoefer-PhysRevA.99.043804}. 
On a physical quantum computer, noise will decrease the magnitude of the coherences at a rate $\Gamma_\mathrm{dec}$. 
Hence, in order to allow us to observe synchronization, the signal must overcome this decoherence, $\Gamma_\mathrm{signal} > \Gamma_\mathrm{dec}$. 
However, this is incompatible with the requirement $\Gamma_\mathrm{rel} > \Gamma_\mathrm{signal}$ since decoherence is typically stronger than energy relaxation, $\Gamma_\mathrm{dec} > \Gamma_\mathrm{relax}$. 
Therefore, to study synchronization on a physical quantum computer, the natural energy relaxation rate $\Gamma_\mathrm{rel}$ must be artificially increased.

This can be achieved by the following circuit, also shown in Fig.~\ref{fig:Mapping}(d).
\begin{align}
	\Qcircuit @C=0.5em @R=0.3em @!R {
	\lstick{q} & 
		\qw & 
		\qw & 
		\ctrl{1} & 
		\targ & 
		\qw & 
		\qw & 
		\qw \\
	 \lstick{a} & 
		\qw & 
		\qw & 
		\gate{U_3(\theta,0,0)} & 
		\ctrl{-1} & 
		\meter & 
		\push{\rule{.6em}{0em}\ket{0}\rule{.2em}{0em}} \qw & 
		\qw \\
	 }
	 \label{eqn:SM:LCStabilization:Circuit1}
\end{align}
This circuit maps an initial state $\ket{\psi_0}_q \otimes \ket{0}_a = (\alpha \ket{0}_q + \beta \ket{1}_q) \otimes \ket{0}_a$ to the state 
\begin{align*}
	\left[ \alpha \ket{0}_q + \beta \cos \left( \frac{\theta}{2} \right) \ket{1}_q \right] \otimes \ket{0}_a + \beta \sin \left( \frac{\theta}{2} \right) \ket{0}_q \otimes \ket{1}_a 
\end{align*}
immediately before the measurement. 
If we set $\sin^2 (\theta/2) = \Gamma \d t \ll 1$, the measurement projects the state of qubit $q$ to $\ket{\psi_{\d t}}\vert_1 = \ket{0}_q$ at a probability $\Gamma \abs{\beta}^2 \d t$, or to
\begin{align*}
	\ket{\psi_{\d t}}\vert_0 
	&= \alpha \left( 1 + \frac{\Gamma}{2} \abs{\beta}^2 \d t \right) \ket{0}_q \\
	&+ \beta \left( 1 - \frac{\Gamma}{2} \d t + \frac{\Gamma}{2} \abs{\beta}^2 \d t \right) \ket{1}_q + \mathcal{O}(\d t^2)
\end{align*}
at a probability $1 - \Gamma \abs{\beta}^2 \d t$. 
This is precisely the evolution of the state vector $\ket{\psi}$ in a stochastic Schr\"odinger equation of the form
\begin{align}
	\d \ket{\psi} 
	&= \left[ -i \left( -i \frac{\Gamma}{2} \sigma_- \sigma_+ \right) + \frac{\Gamma}{2} \bra{\psi} \sigma_- \sigma_+ \ket{\psi} \right] \ket{\psi} \d t \nonumber \\
	&+ \left[ \frac{\sigma_+ \ket{\psi}}{\sqrt{\bra{\psi} \sigma_- \sigma_+ \ket{\psi}}} - \ket{\psi} \right] \d N \comma
	\label{eqn:SM:LCStabilization:SSE}
\end{align}
where $\d N \in \{0,1\}$ is a stochastic Poissonian increment with expectation value $\mathrm{E}(\d N) = \Gamma \bra{\psi} \sigma_- \sigma_+ \ket{\psi} \d t = \abs{\beta}^2 \Gamma \d t$ \cite{WisemanMilburn}. 
The unconditional quantum master equation for the density matrix $\hat{\rho} = \mathrm{E}[\ket{\psi} \bra{\psi}]$ corresponding to Eq.~\eqref{eqn:SM:LCStabilization:SSE} describes single-qubit relaxation,
\begin{align}
	\frac{\d}{\d t} \hat{\rho} = \Gamma \mathcal{D}[\hat{\sigma}_+] \hat{\rho} \fullstop
	\label{eqn:SM:LCStabilization:QMEtoSSE}
\end{align}
Note that we are using the quantum-information definition of the single-qubit basis states, \ie, $\hat{\sigma}_z \ket{0} = + \ket{0}$ and $\hat{\sigma}_z \ket{1} = - \ket{1}$. 
Therefore, Eq.~\eqref{eqn:SM:LCStabilization:QMEtoSSE} actually describes relaxation since $\hat{\sigma}_+ \ket{1} = \ket{0}$.

Thus, a measurement result of $1$ on the ancillary qubit $a$ represents the release of an excitation from the qubit $q$ into the environment and resets the qubit $q$ to its ground state.

A controlled unitary gate is implemented by at least two CNOT operations \cite{Barenco-PhysRevA.52.3457,Vatan-PhysRevA.69.032315}.
Thus, the circuit given above requires at least three CNOT operations. 
An alternative circuit which performs exactly the same transformation of the initial state $\ket{\psi_0}_q \otimes \ket{0}_a$, but requires only two CNOT gates is the following.
\begin{widetext}
\begin{align}
	\Qcircuit @C=0.5em @R=0.3em @!R {
	\lstick{q} & 
		\gate{U_2(-\pi,0)} & 
		\targ & 
		\gate{U_2(-\frac{\pi}{2},0)} & 
		\targ & 
		\gate{U_1(-\frac{\pi}{2})} & 
		\qw &
		\qw &
		\qw \\
	 \lstick{a} & 
		\gate{U_3(-\frac{\theta}{2},-\frac{\pi}{2},\pi)} & 
		\ctrl{-1} & 
		\gate{U_3(-\frac{\theta}{2},\pi,\frac{\pi}{2})} & 
		\ctrl{-1} & 
		\gate{U_1(-\frac{\pi}{2})} & 
		\meter & 
		\push{\rule{.6em}{0em}\ket{0}\rule{.2em}{0em}} \qw & 
		\qw \\
	 }
	 \label{eqn:SM:LCStabilization:Circuit2}
\end{align}
\end{widetext}
Despite the fact that both circuits ideally perform the same transformation of an initial state $\ket{\psi_0}_q \otimes \ket{0}_a$, they will perform differently on a NISQ device.
The parameters of the quantum computer fluctuate in time and are recalibrated once a day. 
Therefore, on each day we choose the circuit that induces the least coherences in the limit-cycle state for the given gate errors.

\section{Dissipative limit-cycle stabilization}
\label{sec:App:Stabilization}
Here, we provide additional information on the dissipative stabilization of the limit cycle. 
Figure~\ref{fig:SM:FigA2a} supplements Figs.~\ref{fig:Characterization}(b,c) and shows the coherences $\hat{\rho}_{k,X}$ between the spin-$1$ states and the surplus state $\ket{X}$. 
As discussed in Sec.~\ref{sec:Characterization}, the dissipative stabilization of the limit cycle by the quantum circuits~\eqref{eqn:SM:LCStabilization:Circuit1} and~\eqref{eqn:SM:LCStabilization:Circuit2} induces a small amount of coherence between the spin-$1$ states on an actual NISQ computer. 
The amplitude of this noise is represented by the gray circle.
Figure~\ref{fig:SM:FigA2a}(a) shows that the coherences $\hat{\rho}_{k,X}$ of the limit-cycle state are much smaller than the noise level, \emph{i.e.}, the state $\ket{X}$ is indeed well decoupled from the spin-$1$ subspace.

If an external signal is applied to the limit-cycle oscillator, $\varepsilon > 0$, coherences between the spin-$1$ states and the state $\ket{X}$ are built up.
These nonzero coherences arise because we use a simplified version of the quantum circuit shown in Fig.~\ref{fig:Mapping}(b).
As discussed in Sec.~\ref{sec:Dealing}, we approximated the time evolution of the signal by uncontrolled single-qubit rotations. 
Figure~\ref{fig:SM:FigA2a}(b) justifies this simplification because the coherences $\hat{\rho}_{k,X}$ remain well below the noise amplitude of the limit-cycle coherences, \emph{i.e.}, they represent negligible higher-order corrections to the quantum synchronization dynamics.

\begin{figure*}
	\centering
	\includegraphics[width=\textwidth]{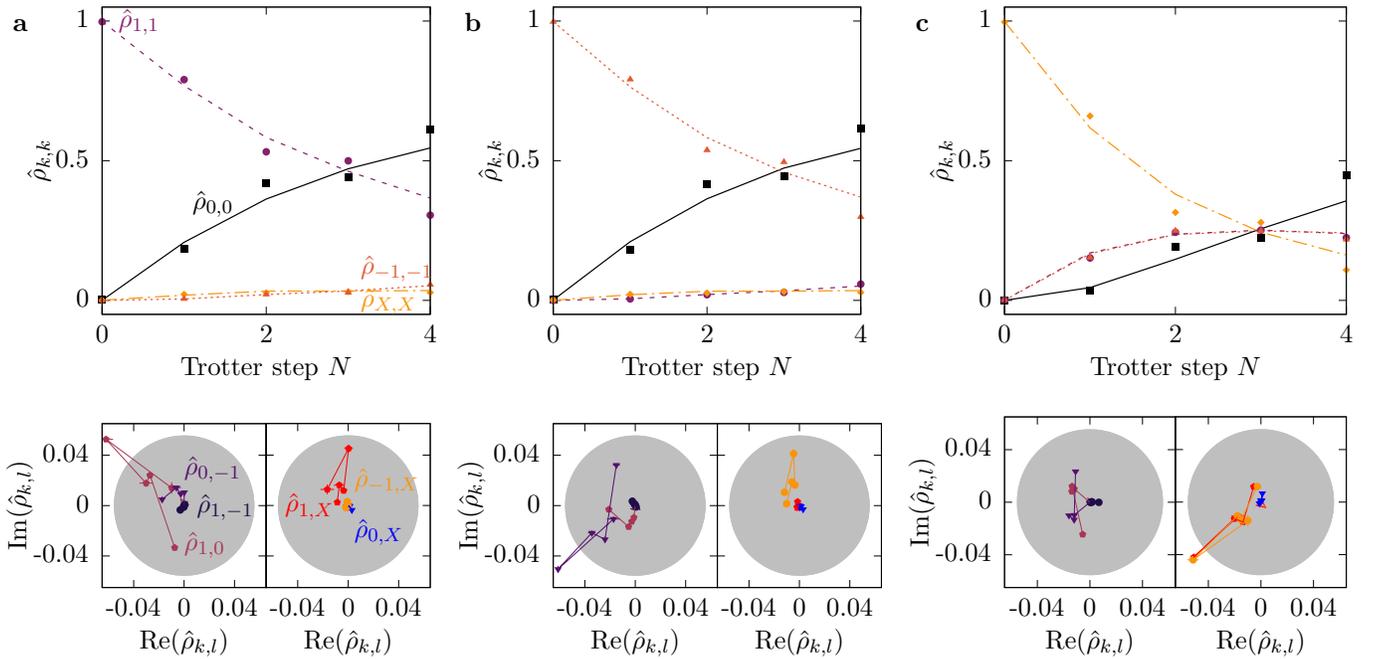}
	\caption{Time evolution of the populations (top row) and the coherences (bottom row) if only the dissipative stabilization mechanism of the limit cycle is switched. on. 
		The initial state is 
		(a) $\ket{+1}$, (b) $\ket{-1}$, and (c) $\ket{X}$.
		Parameters are the same as in Fig.~\ref{fig:Characterization}(b).}
	\label{fig:SM:FigA2b}
\end{figure*}

Finally, Fig.~\ref{fig:SM:FigA2b} shows the evolution of the populations and coherences under the dissipative time evolution if an initial state different from $\ket{0}$ is chosen. 
This data confirms that the dissipative stabilization mechanism given by Eqs.~\eqref{eqn:SM:LCStabilization:Circuit1} and~\eqref{eqn:SM:LCStabilization:Circuit2} transfers population from the initial state to the state $\ket{0}$. 
The coherences stay below the noise level of the limit cycle except for transient dynamics associated with the state transfer.


\begin{thebibliography}{999}

\bibitem{Pikovsky}
	A.\ Pikovsky, M.\ Rosenblum, and J.\ Kurths, 
	\emph{Synchronization. A universal concept in nonlinear sciences}
	(Cambridge University Press, Cambridge, 2001).

\bibitem{Adler-IRE.1697085}
	R.\ Adler, 
 	Proceedings of the IRE, \textbf{34}, 351 (1946).

\bibitem{Pecora-PhysRevLett.64.821}
	L.\ M.\ Pecora and T.\ L.\ Carroll, 
	Phys. Rev. Lett.\ \textbf{64}, 821 (1990). 
	\url{https://link.aps.org/doi/10.1103/PhysRevLett.64.821}

\bibitem{Chagnac-JNeurophysiol.62.1149}
	Y.\ Chagnac-Amitai and B.\ W.\ Connors,
	Journal of Neurophysiology \textbf{62}, 1149 (1989). 

\bibitem{Zhirov-EPJD.38.375}
	O.\ V.\ Zhirov and D.\ L.\ Shepelyansky,
	Eur.\ Phys.\ J.\ D \textbf{38}, 375 (2006).
	\url{https://doi.org/10.1140/epjd/e2006-00011-9}

\bibitem{Nigg-PhysRevA.97.013811}
	S.\ E.\ Nigg, 
	Phys.\ Rev.\ A \textbf{97}, 013811 (2018).
	\url{https://link.aps.org/doi/10.1103/PhysRevA.97.013811}

\bibitem{Ludwig-PhysRevLett.111.073603}
	M.\ Ludwig and F.\ Marquardt, 
	Phys.\ Rev.\ Lett.\ \textbf{111}, 073603 (2013).
	\url{https://link.aps.org/doi/10.1103/PhysRevLett.111.073603}

\bibitem{Walter-PRL.112.094102}
	S.\ Walter, A.\ Nunnenkamp, and C.\ Bruder, 
	Phys.\ Rev.\ Lett.\ \textbf{112}, 094102 (2014).
  	\url{http://link.aps.org/doi/10.1103/PhysRevLett.112.094102}

\bibitem{Lee-PRL.111.234101}
	T.\ E.\ Lee and H.\ R.\ Sadeghpour,
	Phys.\ Rev.\ Lett.\ \textbf{111}, 234101 (2013).
	\url{https://link.aps.org/doi/10.1103/PhysRevLett.111.234101}

\bibitem{Hush-PhysRevA.91.061401}
	M.\ R.\ Hush, W.\ Li, S.\ Genway, I.\ Lesanovsky and A.\ D.\ Armour, 
	Phys.\ Rev.\ A \textbf{91}, 061401(R) (2015). 
	\url{http://dx.doi.org/10.1103/PhysRevA.91.061401}

\bibitem{Holmes-PhysRevE.85.066203}
	C.\ A.\ Holmes, C.\ P.\ Meaney, and G.\ J.\ Milburn, 
	Phys.\ Rev.\ E \textbf{85}, 066203 (2012). 
	\url{https://link.aps.org/doi/10.1103/PhysRevE.85.066203}

\bibitem{Zalalutdinov-ApplPhysLett.83.3281} 
	M.\ Zalalutdinov, K.\ L.\ Aubin, M.\ Pandey, A.\ T.\ Zehnder, R.\ H.\ Rand, H.\ G.\ Craighead, and J.\ M.\ Parpia,
	Appl.\ Phys.\ Lett.\ \textbf{83}, 3281 (2003).
	\url{https://doi.org/10.1063/1.1618363}

\bibitem{Hossein-Zadeh-ApplPhysLett.93.191115} 
	M.\ Hossein-Zadeh and K.\ J.\ Vahala,
	Appl.\ Phys.\ Lett. \textbf{93}, 191115 (2008).
	\url{https://aip.scitation.org/doi/10.1063/1.3028024}

\bibitem{Zhang-PhysRevLett.115.163902}
	M.\ Zhang, S.\ Shah, J.\ Cardenas, and M.\ Lipson,
	Phys.\ Rev.\ Lett. \textbf{115}, 163902 (2015).
	\url{https://link.aps.org/doi/10.1103/PhysRevLett.115.163902}

\bibitem{Bagheri-PhysRevLett.111.213902}
	M.\ Bagheri, M.\ Poot, L.\ Fan, F.\ Marquardt, and H.\ X.\ Tang,
	Phys.\ Rev.\ Lett.\ \textbf{111}, 213902 (2013).
	\url{https://link.aps.org/doi/10.1103/PhysRevLett.111.213902}

\bibitem{Shlomi-PhysRevE.91.032910}
	K.\ Shlomi, D.\ Yuvaraj, I.\ Baskin, O.\ Suchoi, R.\ Winik, and E.\ Buks,
	Phys.\ Rev.\ E \textbf{91}, 032910 (2015).
	\url{https://link.aps.org/doi/10.1103/PhysRevE.91.032910}

\bibitem{Seitner-PhysRevLett.118.254301}
	M.\ J.\ Seitner, M.\ Abdi, A.\ Ridolfo, M.\ J.\ Hartmann, and E.\ M.\ Weig,
	Phys.\ Rev.\ Lett.\ \textbf{118}, 254301 (2017).
	\url{https://link.aps.org/doi/10.1103/PhysRevLett.118.254301}
	
\bibitem{Gil-Santos-PhysRevLett.118.063605}
	E.\ Gil-Santos, M.\ Labousse, C.\ Baker, A.\ Goetschy, W.\ Hease, C.\ Gomez, A.\ Lema\^{\i}tre, G.\ Leo, C.\ Ciuti, and I.\ Favero,
	Phys.\ Rev.\ Lett.\ \textbf{118}, 063605 (2017).
	\url{https://link.aps.org/doi/10.1103/PhysRevLett.118.063605}

\bibitem{Bekker-optica.4.1196}
	C.\ Bekker, R.\ Kalra, C.\ Baker, and W.\ P.\ Bowen,
	Optica \textbf{4}, 1196 (2017).
	\url{http://www.osapublishing.org/optica/abstract.cfm?URI=optica-4-10-1196}

\bibitem{Toth-PhysLettA.382.2233}
	L.\ D.\ Toth, N.\ R.\ Bernier, A.\ K.\ Feofanov, and T.\ J.\ Kippenberg,
	Physics Letters A \textbf{382}, 2233 (2018).
	\url{http://www.sciencedirect.com/science/article/pii/S0375960117302876}

\bibitem{Huang-OptExpr.26.8275}
	K.\ Huang and M.\ Hossein-Zadeh,
	Opt.\ Express \textbf{7}, 8275 (2018).
	\url{http://www.opticsexpress.org/abstract.cfm?URI=oe-26-7-8275}

\bibitem{Roulet-PRL.121.053601}
	A.\ Roulet and C.\ Bruder,
	Phys.\ Rev.\ Lett.\ \textbf{121}, 053601 (2018). 
	\url{https://link.aps.org/doi/10.1103/PhysRevLett.121.053601}

\bibitem{NielsenChuang}
	M.\ A.\ Nielsen and I.\ L.\ Chuang, 
	(Cambridge University Press, Cambridge, 2000).

\bibitem{Lloyd-science.273.1073}
	S.\ Lloyd,
	Science \textbf{273}, 1073 (1996).
	\url{https://science.sciencemag.org/content/273/5278/1073}

\bibitem{IBMQX}
	\url{https://quantum-computing.ibm.com/}

\bibitem{Preskill-Quantum.2.2521}
	J.\ Preskill,
	Quantum \textbf{2}, 79 (2018).
	\url{https://doi.org/10.22331/q-2018-08-06-79}

\bibitem{Peruzzo-ncomms.5.4213}
	A.\ Peruzzo, J.\ McClean, P.\ Shadbolt, M.-H.\ Yung, X.-Q.\ Zhou, P.\ J.\ Love, A.\ Aspuru-Guzik, and J.\ L.\ O'Brien,
	Nature Communications \textbf{5}, 4213 (2014).
	\url{https://www.nature.com/articles/ncomms5213}

\bibitem{Kandala-Nature.549.242}
	A.\ Kandala, A.\ Mezzacapo, K.\ Temme, M.\ Takita, M.\ Brink, J.\ M.\ Chow, and J.\ M.\ Gambetta,
	Nature \textbf{549}, 242 (2017).
	\url{https://www.nature.com/articles/nature23879}

\bibitem{Reiner-QST.4.035005}
	J.-M.\ Reiner, F.\ Wilhelm-Mauch, G.\ Sch\"on, and M.\ Marthaler, 
	Quantum Science and Technology \textbf{4}, 035005 (2019)
	\url{https://doi.org/10.1088%2F2058-9565%2Fab1e85}

\bibitem{Pollmann-1906.06343}
	A.\ Smith, M.\ S.\ Kim, F.\ Pollmann, and J.\ Knolle,
	npj Quantum Inf.\ \textbf{5}, 106 (2019).
	\url{https://www.nature.com/articles/s41534-019-0217-0}

\bibitem{Lloyd-PhysRevA.65.010101}
	S.\ Lloyd and L.\ Viola, 
	Phys.\ Rev.\ A \textbf{65}, 010101(R) (2001).
	\url{https://link.aps.org/doi/10.1103/PhysRevA.65.010101}

\bibitem{Bacon-PhysRevA.64.062302}
	D.\ Bacon, A.\ M.\ Childs,  I.\ L.\ Chuang, J.\ Kempe, D.\ W.\ Leung, and X.\ Zhou
	Phys.\ Rev.\ A \textbf{64}, 062302 (2001).
	\url{https://link.aps.org/doi/10.1103/PhysRevA.64.062302}

\bibitem{Kliesch-PhysRevLett.107.120501}
	M.\ Kliesch, T.\ Barthel, C.\ Gogolin, M.\ Kastoryano, and J.\ Eisert,
	Phys.\ Rev.\ Lett.\ \textbf{107}, 120501 (2011).
	\url{https://link.aps.org/doi/10.1103/PhysRevLett.107.120501}

\bibitem{Garcia-Perez-1906.07099}
	G.\ Garc\'ia-P\'erez, M.\ A.\ C.\ Rossi, and S.\ Maniscalco,
	npj Quantum Inf.\ \textbf{6}, 1 (2020). 
	\url{https://www.nature.com/articles/s41534-019-0235-y}

\bibitem{Koppenhoefer-PhysRevA.99.043804}
	M.\ Koppenh\"ofer and A.\ Roulet,
	Phys.\ Rev.\ A \textbf{99}, 043804 (2019).
	\url{https://link.aps.org/doi/10.1103/PhysRevA.99.043804}

\bibitem{WisemanMilburn}
	H.\ M.\ Wiseman and G.\ J.\ Milburn, 
	\emph{Quantum Measurement and Control}
	(Cambridge University Press, Cambridge, 2010).

\bibitem{IBMHardware}
	A.\ D.\ Corcoles, A.\ Kandala, A.\ Javadi-Abhari, D.\ T.\ McClure, A.\ W.\ Cross, K.\ Temme, P.\ D.\ Nation, M.\ Steffen, and J.\ M.\ Gambetta, 
	arXiv:1910.02894 (2019).
	\url{https://arxiv.org/abs/1910.02894}

\bibitem{Qiskit}
	H.\ Abraham et al., 
	\emph{Qiskit: An Open-source Framework for Quantum Computing}
	(2019). 
	\url{https://www.doi.org/10.5281/zenodo.2562110}

\bibitem{Smolin-PhysRevLett.108.070502}
	J.\ A.\ Smolin, J.\ M.\ Gambetta, and G.\ Smith,
	\emph{Efficient Method for Computing the Maximum-Likelihood Quantum State from Measurements with Additive Gaussian Noise},
	Phys.\ Rev.\ Lett.\ \textbf{108}, 070502 (2012).
	\url{https://link.aps.org/doi/10.1103/PhysRevLett.108.070502}

\bibitem{Qutip}
	J.\ R.\ Johansson, P.\ D.\ Nation, and F.\ Nori, 
	\emph{QuTiP: An open-source Python framework for the dynamics of open quantum systems},
	Computer Physics Communications \textbf{183}, 1760 (2012). 
	\url{https://www.sciencedirect.com/science/article/pii/S0010465512000835}	
	
\bibitem{Wiersig-PhysRevA.93.033809}
	J.\ Wiersig, 
	Phys.\ Rev.\ A \textbf{93}, 033809 (2016).
	\url{https://link.aps.org/doi/10.1103/PhysRevA.93.033809}

\bibitem{Metelmann-PhysRevX.5.021025}
	A.\ Metelmann and A.\ A.\ Clerk,
	Phys.\ Rev.\ X \textbf{5}, 021025 (2015).
	\url{https://link.aps.org/doi/10.1103/PhysRevX.5.021025}

\bibitem{Poyatos-PhysRevLett.77.4728}
	J.\ F.\ Poyatos, J.\ I.\ Cirac, and P.\ Zoller,
	Phys.\ Rev.\ Lett.\ \textbf{77}, 4728 (1996).
	\url{https://link.aps.org/doi/10.1103/PhysRevLett.77.4728}

\bibitem{arxiv.1910.11832}
	A.\ W.\ Laskar, P.\ Adhikary, S.\ Mondal, P.\ Katiyar, S.\ Vinjanampathy, and S.\ Ghosh,
	arXiv:1910.11832 (2019).
	\url{https://arxiv.org/abs/1910.11832}

\bibitem{Barenco-PhysRevA.52.3457}
	A.\ Barenco, C.\ H.\ Bennett, R.\ Cleve, D.\ P.\ DiVincenzo, N.\ Margolus, P.\ Shor, T.\ Sleator, J.\ A. Smolin, and H.\ Weinfurter,
	\emph{Elementary gates for quantum computation},
	Phys.\ Rev.\ A \textbf{52}, 3457 (1995).
	\url{https://link.aps.org/doi/10.1103/PhysRevA.52.3457}

\bibitem{Vatan-PhysRevA.69.032315}
	F.\ Vatan and C.\ Williams,
	\emph{Optimal quantum circuits for general two-qubit gates},
	Phys.\ Rev.\ A \textbf{69}, 032315 (2004).
	\url{https://link.aps.org/doi/10.1103/PhysRevA.69.032315}

\end{thebibliography}
\end{document}